\documentclass[twocolumn,english,aps,prb,showpacs,superscriptaddress]{revtex4}
\usepackage{graphicx}

\begin{document}

\title{Elastic and plastic deformation of graphene, silicene, and boron nitride honeycomb nanoribbons under uniaxial tension: A first-principles density-functional theory study}

\author{M. Topsakal}
\affiliation{UNAM-Institute of Materials Science and
Nanotechnology, Bilkent University, Ankara 06800, Turkey}
\author{S. Ciraci}
\affiliation{UNAM-Institute of Materials Science and
Nanotechnology, Bilkent University, Ankara 06800, Turkey}
\email{ciraci@fen.bilkent.edu.tr} \affiliation{Department of
Physics, Bilkent University Ankara 06800, Turkey}
\date{\today}

\begin{abstract}
This study of elastic and plastic deformation of graphene,
silicene and boron nitride (BN) honeycomb nanoribbons under
uniaxial tension determines their elastic constants and reveals
interesting features. In the course of stretching in the elastic
range, the electronic and magnetic properties can be strongly
modified. In particular, it is shown that the band gap of a
specific armchair nanoribbon is closed under strain and highest
valance and lowest conduction bands are linearized. This way, the
massless Dirac fermion behavior can be attained even in a
semiconducting nanoribbon. Under plastic deformation, the
honeycomb structure changes irreversibly and offers a number of
new structures and functionalities. Cage like structures, even
suspended atomic chains can be derived between two honeycomb
flakes. Present work elaborates on the recent experiments by Jin
\textit{et al.,} Phys. Rev. Lett. \textbf{102}, 205501 (2009)
deriving carbon chains from graphene. Furthermore, the similar
formations of atomic chains from BN and Si nanoribbons are
predicted.
\end{abstract}


\maketitle

\section{introduction}
For last two decades honeycomb structured materials have dominated
nanoscience. The unique orbital symmetry of the honeycomb
structure underlies several exceptional properties of carbon based
nanomaterials, such as fullerenes, nanotubes, graphene and its
quasi one dimensional ribbons. While $\pi$-orbitals are
responsible for the unusual electronic and magnetic properties of
graphene\cite{novoselov,zhang}, its planar flexibility but high
in-plane strength is achieved by $sp^2$-hybrid orbitals. For
example, charge carriers in graphene behave like a massless Dirac
fermions\cite{dirac} due to electron and hole bands showing linear
crossing at the Fermi level. These bands are derived from $\pi$-
and $\pi^*$-states. An unpaired $\pi$-state leads to a local
magnetic moment in a nonmagnetic honeycomb structure\cite{lieb}.
Also the  strong overlap between nearest neighbor $\pi$-orbitals
assures the planar stability of graphene and BN. Silicene, a
honeycomb structure of Si, lacking such an overlap is stabilized
only by puckering\cite{seymur}.

The recent spectroscopy techniques to identify and quantify the
strain profiles in graphene have shown that the Raman peaks shift
considerably under the in-plane strain\cite{mohiuddin} providing a
fundamental tool for graphene-based  micro/nano mechanical
systems. Some theoretical studies\cite{neto,neto-prl} have also
shown that local and uniform strain can be an effective ways of
tuning the electronic structure and transport characteristics of
graphene devices to generate confined states, quantum wires, and
collimation. A related step was given by Kim and
collaborators,\cite{kim} who have  developed a simple method to
grow and transfer large scale, high-quality stretchable graphene
films using CVD on nickel layers, which might enable numerous
applications including in large-scale flexible, stretchable,
foldable transparent electronics.

This study investigates the stretching of quasi one-dimensional
nanoribbons (NRs) of graphene, silicene and BN and predicts that
they attain additional functionalities by changing to a number of
structures with interesting electronic and magnetic properties.
While the elastic deformation with harmonic and anharmonic ranges
and sudden yielding points are common to all NRs, the absence of
sequential elastic deformation stages ending with order-disorder
structural transformation each leading to stepwise necking
constitutes their prime difference from metal
nanowires\cite{agrait,mehrez}. Cage structures of large polygons
can be generated. In particular, atomic chains of C, BN and Si
between honeycomb flakes can form under certain circumstances. The
synthesis of these ultimate one-dimensional atomic chains, which
can form various stable geometries\cite{tongay}, are expected to
be an essential step towards future "active" nanotechnology
applications. We believe that our predictions are relevant for the
current studies investigating the properties of strained
graphene\cite{neto}.

\begin{figure*}
\begin{centering}
\includegraphics[width=17cm]{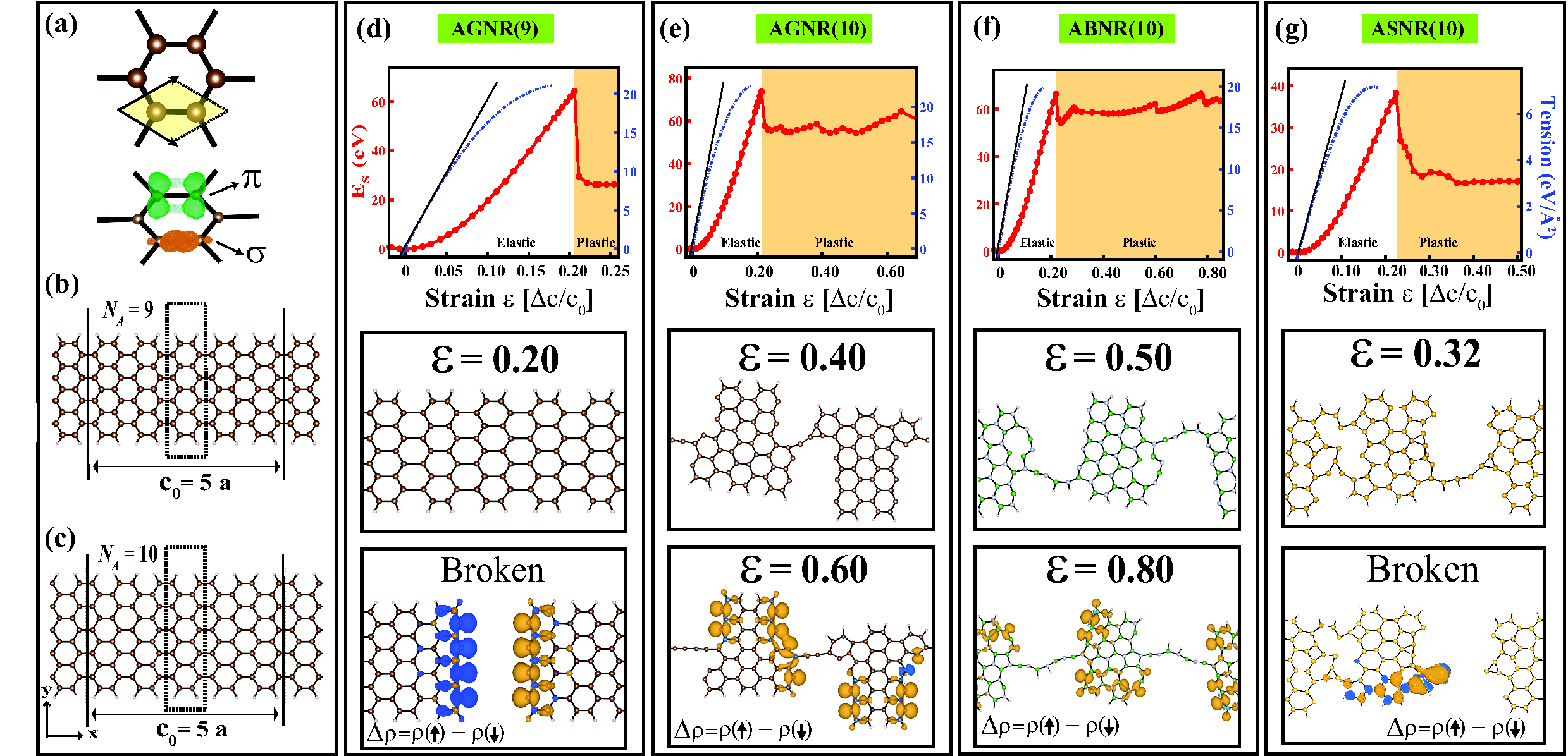}
\par\end{centering}
\caption{(Color online) (a) Two dimensional honeycomb structure
with primitive unit cell. $\pi-$ and $\sigma-$bonds are
schematically described. (b) An armchair graphene nanoribbon
AGNR(9) has mirror symmetry with respect to its axis and $N_A$=9
dimer bonds across the unit cell specifying its width. The unit
cell is delineated by a rectangle with dotted edges. The supercell
consists of five unit cells with $c_{0}=5\times a$. (c) Same for
AGNR($10$), which lacks the mirror symmetry. (d) Response of
AGNR($9$) to uniaxial tension examined in the supercell; variation
of the strain energy $E_{S}$ and tension force $F_{T}$ with shaded
region indicating the plastic range; atomic structure for
$\epsilon=$0.20 just before breaking. Isosurfaces for the
difference of spin up (blue/dark) and spin down (yellow/light)
charge densities, $\Delta\rho=\rho(\uparrow)-\rho(\downarrow)$
calculated for broken pieces show antiferromagnetic order of
zigzag edge states.(c) Same for AGNR($10$). Monatomic carbon
chains connect magnetic pieces. (d) Armchair boron-nitride NR. (e)
Armchair silicene NR.}
\label{fig:Figure-1}
\end{figure*}

\section{Model and Methodology}

We have performed first-principles plane wave calculations within
Density Functional Theory (DFT) using PAW potentials.\cite{paw}
The exchange correlation potential has been approximated by
Generalized Gradient Approximation (GGA) using PW91 \cite{pw91}
functional both for spin-polarized and spin-unpolarized cases.
Recently, spin-polarized calculations within DFT have been carried
out successfully to investigate magnetic properties of vacancy
defects in 2D honeycomb structures. Also interesting spintronic
properties of nanoribbons have been revealed using spin-polarized
DFT\cite{gribbon2}. The success of spin-polarized DFT calculations
has been discussed in Ref{\onlinecite{zeller}.

All structures have been treated within supercell geometry using
the periodic boundary conditions. A plane-wave basis set with
kinetic energy cutoff of 400 eV has been used. The interaction
between monolayers in adjacent supercells is examined as a
function of their spacing. Since the total energy per cell has
changed less than 1 meV upon increasing the spacing from 10 \AA~
to 15 \AA~, we used the spacing of $\sim10$ \AA~ in the
calculations.  In the self-consistent potential and total energy
calculations the Brillouin zone (BZ) is sampled in
\textbf{k}-space using Monkhorst-Pack scheme by 25x1x1 for
nanoribbons. This sampling is scaled according to the size of
superlattices. All atomic positions and lattice constants are
optimized by using the conjugate gradient method, where the total
energy and atomic forces are minimized. The convergence for energy
is chosen as 10$^{-5}$ eV between two steps. Numerical plane wave
calculations are performed by using VASP
package.\cite{vasp1,vasp2}

\section{Elastic and plastic deformation of nanoribbons}

Nanomechanics of both armchair and zigzag NRs of graphene,
silicene and BN  is explored by calculating the mechanical
properties as a response to stretching along the axis of the
ribbon. Mechanical properties are revealed from the strain energy
$E_{S}=E_{T}(\epsilon)-E_{T}(\epsilon=0)$; namely, the total
energy at a given axial strain $\epsilon$ minus the total energy
at zero strain. Segments of quasi-one dimensional NRs are treated
within supercell geometry using periodic boundary conditions. Each
supercell contains $n$ unit cells of the ribbon and hence has the
lattice constant along the axis of ribbon, $c_{0}=na$, $a$ being
lattice constant for the primitive unit cell of NR. The strain,
$\epsilon=\Delta c/c_{0}$, corresponds to a stretching, where the
lattice constant of the strained supercell equals $c=c_{0}+\Delta
c$. The stretching of the ribbon is achieved by first increasing
the optimized lattice constant in increments of
$\Delta\epsilon=$0.01 (namely $c \rightarrow c+\Delta\epsilon
\times c_{0}$) and by uniformly expanding the atomic structure
obtained from previous optimization. Subsequently the atomic
structure is reoptimized keeping the increased lattice constant
$c$ fixed and the corresponding strain energy is calculated. This
process is repeated after each increment of $\Delta\epsilon$. Then
the tension force, $F_{T}=-\partial E_{S}(\epsilon)/\partial c$
and the force constant $\kappa=\partial^{2}E_{S}/\partial c^{2}$
are obtained from the strain energy. Owing to ambiguities in
defining the Young's modulus of 2D honeycomb structures, one can
use in-plane stiffness,
$C=(1/A_{0})\cdot(\partial^{2}E_{S}/\partial\epsilon^{2})$\cite{yakobson1,reddy}.
Here $A_{0}$ is the equilibrium area of the supercell. The
in-plane stiffness can be deduced from $\kappa$ by defining an
effective width for the ribbon.

\begin{figure}
\begin{centering}
\includegraphics[width=8.0cm]{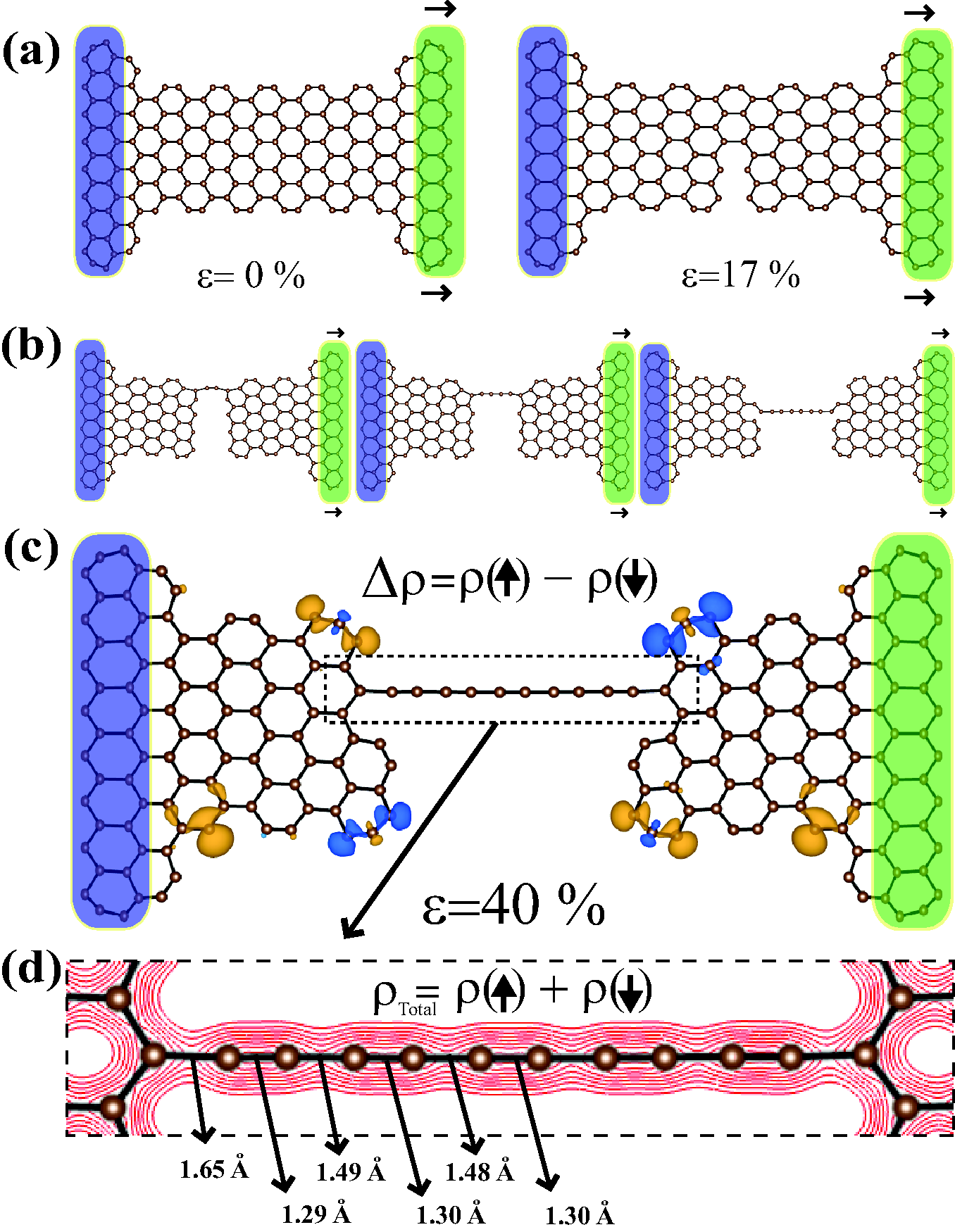}
\par\end{centering}

\caption{(Color online) Stretching of a finite size segment of the
bare armchair graphene NR ($N_{A}$=10) between two tapered ends.
(a) The atomic structure of the NR corresponding to $\epsilon$=0
and $\epsilon$=0.17. Note that a tear starts at one edge of the
stretched NR. (b) In the course of stretching, a hexagon connected
to the chain is transformed to a pentagon by yielding a single
carbon atom to the chain, which, in turn, becomes longer. In the
left panel, one atom is incorporated in the chain when one bond of
the hexagon is broken (c) A suspended chain comprising 12 carbon
atoms is derived from graphene NR at $\epsilon$=0.40. The
difference charge density of different spins states,
$\Delta\rho(\uparrow,\downarrow)$ is also shown. (d) The chain
structure with alternating short and long bonds are highlighted.
The total charge density $\rho_{T}$ of both spins is denser around
short C-C bonds.}

\centering{}\label{fig:Figure-2}
\end{figure}

The variation of strain energy, tensile force and the
corresponding atomic structure of selected armchair NRs are
presented in Fig.~\ref{fig:Figure-1} as a function of $\epsilon$.
Here the bonds parallel to the direction of applied tension is
stretched more than those in other directions. As a result, the
hexagonal symmetry is disturbed, but overall honeycomb like
structure is maintained. The elastic deformation is reversible and
stretched ribbons can return to their original geometry when the
tension is released. In the harmonic range, the force constant is
calculated to be $\kappa$=176, 30 and 144 N/m, for armchair
graphene, silicene and BN NRs having $N_{A}$=10, respectively.
Similarly, the calculated in-plane stiffness for the same ribbons
are, respectively, $C$=292, 51 and 239 N/M. Notably, $\kappa$ and
$C$ values of silicene are lowest due to $sp^2$ orbitals
dehybridized as a result of puckering. Cohesive energies of
graphene, silicene and BN nanoribbons with $N_{A}$=10 (namely
19.77, 10.76, 18.80 eV per atom pair, respectively) show trends
similar to the corresponding in-plane stiffness values. We also
calculated the in-plane stiffness values of 2D graphene, silicene
and BN honeycomb structures to be, respectively, $C$=335 N/m (the
reported experimental value\cite{gr_exp} for graphene:
$C$=340$\pm$50), 62, 258 N/m. Because of edge effects of NRs,
their stiffness values are relatively smaller than those of 2D
structures.

The elastic range ends at the yielding point with the
corresponding critical strain, , where the strain energy drops
suddenly. It should be noted that $\epsilon_{Y}$ and plastic
deformation of nanoribbons are expected to depend on the ambient
temperature, unit cell size, crystalline defects such as vacancy
and time rate of change of stretching. The stochastic nature of
deformation is avoided to some extent by carrying out a slow or
"adiabatic" stretching as explained above, whereby the ribbon is
stretched in small increments and the structure is optimized after
each increments in elastic and plastic ranges. The stretching of
nanoribbon is carried out at 0$^{o}$ K for three different values
of $\Delta\epsilon$, namely 0.05, 0.01, 0.002. The value of
$\epsilon_{Y}$ did not changed for $\Delta\epsilon$=0.01 and
0.002. We therefore concluded that already $\Delta\epsilon$=0.01
corresponds to a very slow or "adiabatic" stretching and reveals
the bare response of nanoribbon without the effect of time rate of
change of stretching. The effect of temperature is investigated by
performing ab-initio molecular dynamic calculations (lasting 2 ps
with time steps of 2x10$^{-15}$ seconds) for $N_A$=10. Results
indicate that $\epsilon_{Y}$ decreases with increasing
temperature. Hence, owing to the softening of acoustical phonons,
$\epsilon_{Y}$=0.22 corresponding to T=0 K is reduced to 0.16 at
T=600 K. We also found that the results are converged if the
supercell size $n \geq 5$. The presence of a vacancy defect in the
ribbon speeds up the yielding by decreasing the value of
$\epsilon_{Y}$.

The plastic deformation stage following the yielding point is the
crucial part in the stretching of NRs having honeycomb structure
and hence is the focus of this work. The response of the ribbon to
the strain after the yielding point is material and geometry
specific. Having determined various effects, which possibly change
the value of $\epsilon_{Y}$, we examine the structure in the range
of plastic (irreversible) deformation through "adiabatic"
stretching. The armchair graphene NR with $N_{A}$=9, i.e.
AGNR($9$) has a mirror symmetry relative to its $x$-axis. This NR
is broken into graphene patches having equilibrium honeycomb
structure just after $\epsilon_{Y}\cong$0.21 in
Fig.~\ref{fig:Figure-1}(a). Whereas the behavior of AGNR($10$)
(which lacks the mirror symmetry) is dramatically different. The
ribbon is torn into two pieces (patches), which are connected by
an atomic chain. In the plastic range, the strain energy $E_{S}$
increases slightly with strain, but eventually decreases each time
when a $sp^{2}$ C-C bond of the zigzag edge of the patch is broken
and a C atom is incorporated into the chain from the graphene
patch. This important result actually predicts the recent finding
by Iijima and his collaborators, who derived monatomic carbon
chain from graphene\cite{iijima}. Carbon atomic chains identified
as cumulene (having double bonds) or polyyne (consisting of
alternating triple and single bonds) have been studied
earlier\cite{saito,tongay,senger}. The chain structure was found
to be stable and linear owing to the strong overlap of
$\pi$-orbitals between adjacent atoms. While infinite chain is
subject to a Peierls distortion, bond alternation and bond length
variation depends on the number of carbon atoms in a finite carbon
chain. The character of the covalent $sp$+$\pi$-bonding between
carbon atoms underlies their unusual chemical, mechanical and
quantum transport properties.

Our results show that the tearing of the armchair graphene NR and
hence the formation of a carbon chain is promoted by a vacancy
defect. Even more surprising is that not only graphene, but also
puckered silicene and flat BN armchair ribbons are plastically
deformed to form atomic chains between patches. Since BN honeycomb
structure is already synthesized, the present results concerning
the formation of BN chains is important and hence is yet to be
realized experimentally.  In contrast to graphene, pieces
(patches) torn from the BN and silicene ribbons allow also
different types of polygons ranging from trigon to heptagon. In
particular, a large hole in the silicene pieces is reminiscent of
the cage structure as if a 2D analogue of metal-organic frameworks
(MOFs). We also note that nonmagnetic armchair NRs attain
spin-polarized (magnetic) ground state after they are broken into
small pieces having zigzag edges\cite{lieb}. This is demonstrated
by isosurfaces for the difference of spin up and spin down charge
densities,
$\Delta\rho(\uparrow,\downarrow)=\rho(\uparrow)-\rho(\downarrow)$
in Fig.~\ref{fig:Figure-1}.

\begin{figure*}
\begin{centering}
\includegraphics[width=16.0cm]{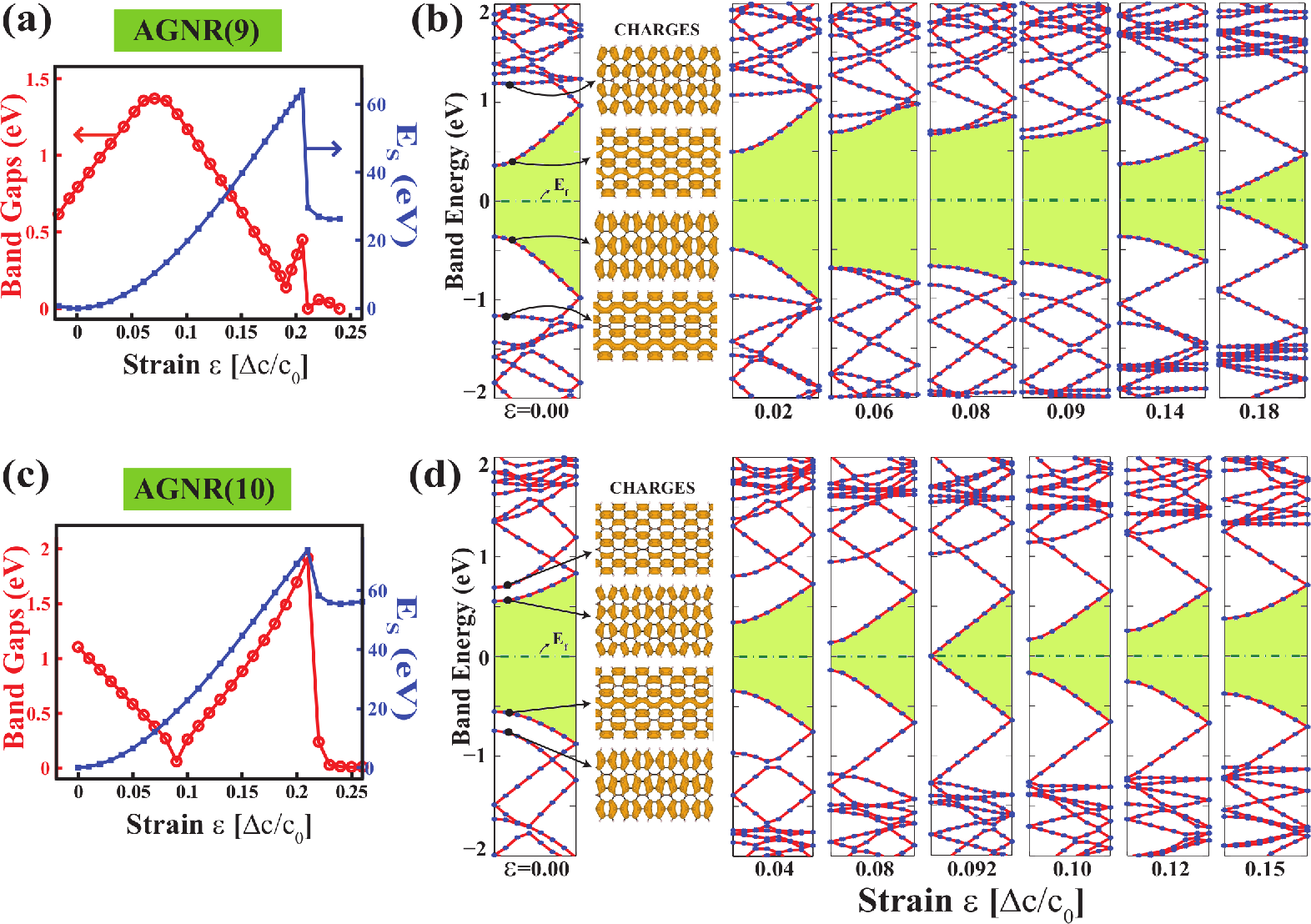}
\par\end{centering}

\caption{(Color online) Variation of the energy band gaps of
AGNR($9$) and AGNR($10$) with the strain from $\epsilon$=0.0 to
$\epsilon$=0.25. (a) The band gap of AGNR($9$) first increases
with increasing strain in the elastic range, passes through a
maximum, then decreases and eventually vanishes after the yielding
point. (b) The band structures for different strain values and the
isosurfaces of charge densities of lowest two conduction and
highest two valance bands for zero-strain configuration. (c)
Variation of the band gap of AGNR($10$) with strain displays a
reverse trend relative to AGNR($9$). (d) same as (b) for
AGNR($10$).}

\centering{}\label{fig:Figure-3}
\end{figure*}

The behavior of a finite segment of bare armchair graphene NR
under uniaxial tension between its two ends is also presented in
Fig.~\ref{fig:Figure-2}(a). The tear, which initiates at one edge,
propagates until the other edge and eventually the chain formation
sets in. Usually triangles of atoms are formed at the region of
junction of the nanoribbon pieces and the atomic chain. Upon
stretching, the apex atom of the triangle is incorporated in the
chain leaving behind a pentagonal (or broken hexagonal) ring as
shown in Fig.~\ref{fig:Figure-2}(b). This way, a carbon atomic
chain with alternating long and short bonds is suspended between
two graphene pieces and grows by sequential implementation of
atoms from these pieces to the chain. Here we point out an
important difference between planar honeycomb NRs and metal
nanorods both stretching in the plastic range. Experimental
studies\cite{agrait} and theoretical simulations\cite{mehrez} have
demonstrated that metallic nanorods elongate in terms of
sequential elastic and yielding stages. At the end of each elastic
stage, the nanorod undergoes an order-disorder transformation
through the yielding stage resulting in a necking. Thus the onset
of a subsequent elastic stage progresses through a smaller cross
section. These structural changes were revealed by in-situ
measurements of two terminal conductance\cite{agrait}. As for NRs
here, the order-disorder transformation occurring at the end of
each elastic stage and leading to necking is absent. Necking of a
NR takes place through tearing. The transmission coefficient,
{\textbf{\textit{T}}} of carbon chain, which is calculated
self-consistently using non-equilibrium Green's Function method
reflects the combined electronic structure of central region and
connecting electrodes. Results to be published elsewhere indicate
that the implementation of a single carbon atom from hexagon to
the chain is reflected to the variation of {\textbf{\textit{T}}}
under constant bias.

\begin{figure}
\begin{centering}
\includegraphics[width=8.0cm]{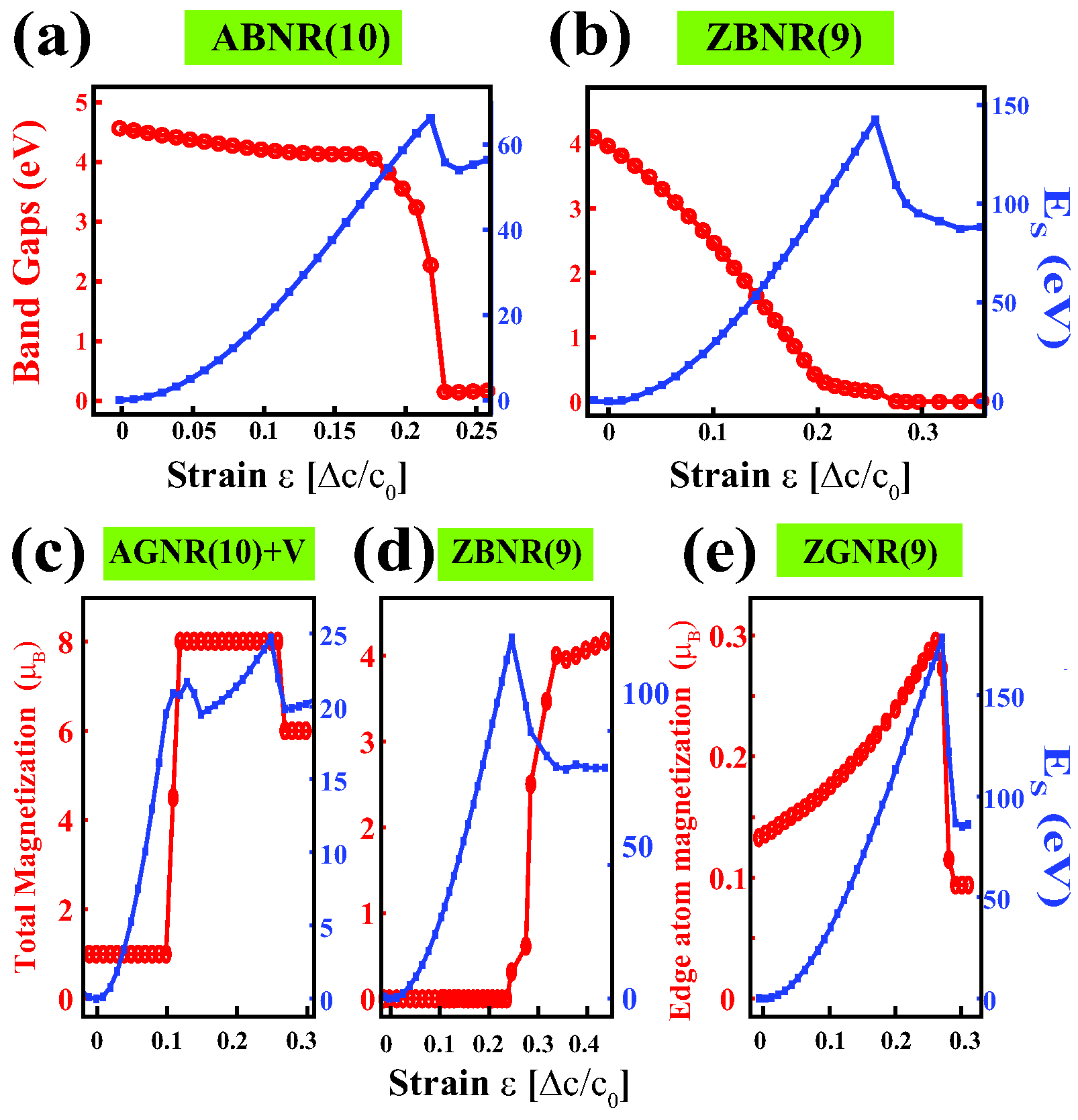}
\par\end{centering}

\caption{(Color online) (a) Variation of the band gap of the
armchair BN nanoribbon, ABNR($10$), with strain $\epsilon$. (b)
Zigzag BN nanoribbon ZBNR($10$). (c) The magnetic moment of
AGNR($10$) having a single vacancy. $\mu$ jumps from 1 $\mu_{B}$
to 8 $\mu_{B}$ after the NR is torn suddenly from one edge. (d)
Antiferromagnetic ZBNR($9$) attains spin-polarized ground state
after the yielding point. (e) Magnetic moment of a single edge
atom of ZGNR($9$) (which is antiferromagnetic in equilibrium)
increases with strain in the elastic range and then falls suddenly
at the yielding point.}

\centering{}\label{fig:Figure-4}
\end{figure}

\section{Variation of electronic and magnetic properties with
strain}

After massive structural changes taking place for $\epsilon >
\epsilon_{Y},$ a NR attains new properties which are absent in its
equilibrium state. For example, the ribbon achieves a higher
chemical reactivity, because of the unsaturated bonds protruding
from atoms having lower coordination. Cumulene by itself is very
reactive. Not only mechanical properties and atomic configuration,
but also the electronic and magnetic properties of nanoribbons can
be modified through stretching as illustrated in
Fig.~\ref{fig:Figure-3}. Depending on their widths, symmetries and
materials the band gaps of nanoribbons exhibit significant
variations in the elastic deformation range, but usually they
vanish in the plastic range. For example, variations of the band
gaps of hydrogen saturated AGNR($9$) and AGNR($10$) with
$\epsilon$ in the elastic range are rather different. The band gap
of AGNR($9$) is $\sim$0.7 eV at $\epsilon$=0.0, but it increases
with increasing strain up to $\epsilon$=0.07, but passes through a
maximum and subsequently decreases to vanish at $\epsilon_{Y}$ as
seen from Fig.~\ref{fig:Figure-3}(a). The band structure near the
band gap and isosurfaces of charge densities of the lowest
(highest) two states in the conduction (valence) band are
presented in Fig.~\ref{fig:Figure-3}(b) for $\epsilon=0.0$.
According to the electronic energy bands calculated as a function
strain, while the highest valence band is lowered, the lowest
conduction band is raised with increasing $\epsilon$ This,
normally, increases the band gap. On the contrary, second highest
(lowest) valence (conduction) band is raised (lowered) with
increasing strain $\epsilon$. Consequently, the band gap of
AGNR($9$) first increases up to $\epsilon$=0.07, but decreases for
$\epsilon > $0.07, where the second bands cross the first ones and
dip in the gap. At the end the orderings of the first and second
valence and conduction bands are switched. In
Fig.~\ref{fig:Figure-3}(c)-(d), the character and orbital
compositions of the first and second valence and conduction bands
are reversed in AGNR($10$). Consequently, the variation of band
gap with the strain is reversed in AGNR($10$) relative to
AGNR($9$). While the highest (lowest) valence (conduction) band is
raised (lowered), second valence (conduction) band is lowered
(raised) and their dispersions is decreased with increasing
$\epsilon$. This way the band gap decreases and is eventually
closed within 3 meV at $\epsilon \sim$ 0.09. Even more remarkable
is that the first conduction and valence bands, which are closed
at \textbf{k}$\rightarrow$ 0, are linearized for \textbf{k}> 0.
This means that the band energy has linear dispersion even for a
small bias $\Delta V$; namely,

\begin{equation}
E({\bf k})- (E_F \pm \Delta V)=\textit{C}{\bf k},
\end{equation}

where \textit{C} is a constant. Accordingly, hydrogen terminated
armchair nanoribbons AGNR($10$), which is normally a
semiconductor, behave like a 2D graphene and hence have carriers,
i.e. holes or electrons, with massless Dirac fermion character at
a specific value of $\epsilon$. This result is somehow unexpected
but has important consequences: The massless Dirac fermion
behavior in 2D graphene originating from linear band crossing at
K-points at the Fermi level disappears when its size is finite.
For example, massless Dirac fermion behavior is absent in an
armchair nanoribbon, which is a nonmagnetic semiconductor at
$\epsilon$=0.0. Whereas the massless Dirac fermion behavior would
be highly desirable to achieve high carrier mobility in these
nanoribbons. Realization of linearized bands as predicted in
Fig.~\ref{fig:Figure-3}(b) is encouraging for various electronic
applications, especially for fast nanoelectronics. We also note
that different response of different bands to the tensile strain
gives rise to a metal-insulator transition. For $\epsilon
>$ 0.09, where the character and orbital composition of the
highest valence and lowest conduction band switches the band gap
starts to open and to increase with increasing tensile strain, but
vanishes suddenly at $\epsilon \sim \epsilon_Y$

Because of zone folding the character and the orbital composition
of the bands at the edges of valance and conduction band are
switched by going from AGNR($9$) to AGNR($10$). Therefore, similar
effects can be observed in other families of AGNR having different
widths, $N_{A}$. On the other hand, different response of
different bands to the strain is an interplay between bond length
and bonding (antibonding) bond energies. In fact, we distinguish
two different bonds; namely lateral (parallel to the ribbon axis)
and tilted bonds, which have different response to $\epsilon$. As
$\epsilon$ in the elastic range increases, the length of the
tilted bonds in AGNR($9$) first increase, passes a maximum and
decreases reminiscent of the variation of the band gap in
Fig.~\ref{fig:Figure-3}(a). The values of the strain corresponding
to maximum of band gap and maximum length of tilted bonds are
close. The elongation of the tilted bonds differ depending on
their location relative to the edge. As for the the lengths of the
lateral bonds, they increase with increasing $\epsilon$, even if
their elongations differ depending on their positions relative to
the edge of the ribbon.

Fig.~\ref{fig:Figure-4} shows the variation of the band gaps and
magnetic properties for other nanoribbons. In contrast to
AGNR($10$), the wide band gap of ABNR($10$) in
Fig.~\ref{fig:Figure-4}(a) doesn't change considerably for
$0<\epsilon<0.20$ in the elastic range. However, the wide band gap
of ZBNR($9$) decreases steadily from 4 to 0.3 eV between
$0<\epsilon<0.20$ as shown in  Fig.~\ref{fig:Figure-4}(b). As a
result of plastic deformation either a spin polarized state is
induced or the existing magnetic state is modified. The magnetic
moment of 1 $\mu_{B}$ of AGNR($10$) having a single carbon vacancy
in a supercell of 5 unit cells increases to $8\mu_{B}$ in the
plastic range as shown in Fig.~\ref{fig:Figure-4}(c). This
dramatic increase of the magnetic moment is attained by the severe
modification of the honeycomb structure after the yielding point.
Figure~\ref{fig:Figure-4}(d)shows that the antiferromangetic spin
state of ZBNR($9$) changes into a ferromagnetic state. The
magnetization of edge atoms of zigzag graphene nanoribbon
ZGNR($9$) increases in the elastic range, but falls to a lower
value as shown Fig.~\ref{fig:Figure-4}(e). These results suggest
that uniaxial strain can be used to monitor the electronic and
magnetic properties of 1D nanoribbons both in the elastic and
plastic deformation ranges.

\section{Conclusions}

In this study the elastic constants of graphene, silicene and BN
zigzag and armchair nanoribbons are determined their unusual
featurs are revealed under tensile stress. Their atomic,
electronic and magnetic structures are examined under elastic and
plastic deformation range attained by "adiabatic" stretching. We
found that in the course of elastic stretching the electronic
structure of these nanoribbons are strongly modified. The
variation of band gap is sample and materials specific. For
example, we showed that the variations of the band gaps of
AGNR($9$) and AGNR($10$) with strain display reverse trends. The
variation of band gap involves a complex interplay of zone
folding, diverse elongation of lateral and tilted C-C bonds, and
different orbital composition of the first and second valance and
conduction bands. In particular, the closing of gap and
linearization of highest and lowest conduction bands of a hydrogen
saturated, armchair nanoribbon may have important implications ,
as such that massless Dirac fermion character can be realized even
in semiconducting armchair nanoribbons. Unusual responses of band
gaps to the strain are also obtained in different types
nanoribbons.

The ending of elastic range and the onset of plastic deformation
leading to diverse structural deformations and magnetic states in
periodic and finite size nanoribbons is another interesting
outcome of this study. Structures having large holes are
reminiscent of a metal-organic frameworks, MOFs. We showed that
long monatomic carbon chains can form in the course of stretching.
Our prediction, that suspended atomic chains can also be derived
from BN and silicene nanoribbons under stretching in the plastic
range is yet to be realized experimentally.

\section{Acknowledgement}

Computing resources were partly provided by the National Center
for High Performance Computing of Turkey (UYBHM) under grant
number 2-024-2007. This work is partially supported by TUBA,
Academy of Science of Turkey.

\end{document}